\documentclass[12pt]{article}
\usepackage[right=1.25in,left=1.25in,top=1.1in,bottom=1.1in]{geometry}
\usepackage{hyperref}
\hypersetup{colorlinks, citecolor=blue, filecolor=blue, linkcolor=blue, urlcolor=blue}
\usepackage{graphicx}
\usepackage{url}
\usepackage{amsmath,amsthm} 
\usepackage{engord}
\usepackage{float}
\usepackage{subfig}
\usepackage{pdflscape}
\usepackage{booktabs}
\usepackage{pgfplots}
\usepackage[linesnumbered,ruled,vlined]{algorithm2e}
\usepackage{booktabs}
\usepackage{lipsum}  
\pgfplotsset{compat=1.14}
\pgfplotsset{every axis label/.append style={font=\tiny}}
\usepackage[labelsep=period]{caption} 

\usepackage{amssymb} 
\usepackage{multirow}

\usepackage{xr}

\usepackage{setspace}
\usepackage{sectsty}
\sectionfont{\large}
\subsectionfont{\normalsize}
\subsubsectionfont{\normalsize}

\title{ \vspace*{-2.5cm} \hspace*{-0.5cm} Approximating Auction Equilibria with Reinforcement Learning\footnote{This paper benefitted from advice and feedback from Prof John Rust, Prof Nathan Miller and Prof Harry Paarsch and the seminar participants of Georgetown's EconBrew seminars.}}

\author{Pranjal Rawat\thanks{PhD Candidate in Economics, Georgetown University
\href{mailto:pp712@georgetown.edu}{pp712@georgetown.edu}}}

\date{ \vspace*{0.5cm} \today\\
} 

\begin{document}

\bgroup
\let\footnoterule\relax

\begin{singlespace}
\maketitle

\begin{abstract}
    \noindent Traditional methods for computing equilibria in auctions become computationally intractable as auction complexity increases, particularly in multi-item and dynamic auctions. This paper introduces a self-play based reinforcement learning approach that employs advanced algorithms such as Proximal Policy Optimization and Neural Fictitious Self-Play to approximate Bayes-Nash equilibria. This framework allows for continuous action spaces, high-dimensional information states, and delayed payoffs. Through self-play, these algorithms can learn robust and near-optimal bidding strategies in auctions with known equilibria, including those with symmetric and asymmetric valuations, private and interdependent values, and multi-round auctions. 
\end{abstract}
\end{singlespace}
\thispagestyle{empty}

\clearpage
\egroup
\setcounter{page}{1}


\section{Introduction\label{sec
}}

\noindent Auctions are a time-tested way for sellers to discover prices and allocate items to those who value them the most. As marketplaces and platforms become increasingly virtual, there is an even greater scope to design markets to improve efficiency, revenue, fairness, etc. Game-theoretic analysis has shed much light on strategic behavior in simple auctions with few bidders and simple rules. These insights have been successfully applied to spectrum, procurement, and resale auctions (Milgrom 2004).

However, game-theoretic analysis has struggled to make inroads when auctions begin to get complex: many bidders, multiple items, multiple rounds, varying information disclosures, and payment rules. As we move beyond the textbook cases, it quickly becomes apparent that solving these auctions is highly challenging. The few equilibria that we know in these settings are obtained through very clever arguments that simplify the problem considerably through symmetry and capture extreme cases (Wilson 1987, Cho et al. 2024).

Auctions are typically modeled as Bayesian games of incomplete information. Conitzer and Sandholm (2008) show that computing pure equilibria in 2x2 auctions is NP-complete, i.e., it is computationally infeasible for any algorithm to guarantee a solution in polynomial time. While we know that mixed equilibria must exist (Nash 1951), finding such equilibria is PPAD-hard (Daskalakis, Goldberg, and Papadimitriou 2009). Cai and Papadimitriou (2014) demonstrated the PP-hard complexity of finding exact Bayesian Nash equilibria in specific simultaneous auctions. This and other results lead Cai and Papadimitriou (2014) to conclude that \textit{``we now know that the auctions of Vickrey and Myerson are isolated areas of light in a sea of dark"}.

In parallel to these developments, recent advances in AI have led to breakthroughs in solving games of imperfect information. Reinforcement learning is an active area of research and has shown remarkable progress in finding powerful strategies in games like Chess, Go, and Poker. The central ingredient in these cases is a flexible representation of the value of states and an iterative method to obtain feedback and refine policies. The success of reinforcement learning in games of perfect and imperfect information motivates their use in finding theoretical market equilibria.

In this paper, I show that simple but sophisticated policy gradient algorithms can converge to pure Nash equilibrium in simple auctions through self-play. These algorithms use reinforcement learning, or learning through exploration and feedback, to continually refine strategies by playing against themselves. I focus on approximate equilibria, where participants play $\epsilon$-best responses to each other. At the theoretical equilibria, exploitability is zero, and players play a perfect best response to each other. This allows us to check how close approximate equilibria are to the actual equilibria.

The theoretical benchmarks used here rely on classic single-item auction theory, which is rooted in the works of Vickrey (1961), Wilson (1969), Milgrom and Weber (1981, 1982), and Myerson (1981), with an elegant summary in Krishna (2009). Before we try to find equilibria, it is helpful to know if they exist and are unique. Generally, mixed Nash equilibriums exist (Nash 1950, Jackson and Swinkels 2005). As Castro and Karney (2011) survey, pure Nash equilibria are also guaranteed to exist in a wide range of symmetric and asymmetric settings. With private and symmetric values, in the classical auction formats, equilibria are generally unique and symmetric (Maskin and Riley 2003, Lebrun 2006). Equilibria remain unique under asymmetry under mild conditions. Kaplan and Zamir (2011) provide an interesting counterexample. Lizzeri and Persico (2000) show that most two-player auctions have unique equilibria in general settings.

Several numerical methods have been developed to solve auctions. Armantier et al. (2008) use Monte Carlo simulations to compute gradients w.r.t policy parameters and payoffs and use hill-climbing methods to improve policies. Rabinovich et al. (2013) modify fictitious play for finite actions and continuous valuations to solve simultaneous auctions with substitutes and complementarities. The Gambit software (McKelvey and McLennan, 1996; McKelvey et al., 2016) can solve Bayes-Nash for finite actions and valuations. Bosshard et al. (2020) alternate a search step that simplifies the strategy space to find $\epsilon$-BNE and a verification step that checks the equilibria in the entire strategy space. Bichler et al. (2021) develop neural network-based algorithms to solve auctions via self-play. Bichler et al. (2023) rely on online convex optimization techniques. This paper is similar to these but differs in using more straightforward best-response dynamics and conducting a more comprehensive experimental verification. This paper also differs in its focus on continuous smooth policies and multi-round dynamic auctions.

The remainder of the paper is as follows. Section 2 recasts Bayesian games and auctions into the reinforcement learning framework. Section 3 presents the algorithm. Section 4 presents extensive experiments. Section 5 summarizes and concludes.

\section{Bayesian Games}

Bayesian games are characterized by players with incomplete information, typically about other players' types, which are private. Each player's type is drawn from a known probability distribution, and this type influences their strategy. Formally, a Bayesian game is defined as \( G = (I, V, O, A, f, u) \), where \( I \) is the set of players, \( V \) represents players' private types, \( O \) is the set of possible observations, \( A \) is the action space, \( f \) is the probability density function of types and observations, and \( u \) represents the utility functions of the players.

\subsection{Bayes Nash Equilibria}

In Bayesian games, players select actions based on their observations and types. A strategy for player $i$ is a mapping from observations and types to actions, denoted $\beta_i: O_i \times V_i \rightarrow A_i$. The central solution concept is the \emph{Bayesian Nash Equilibrium} (BNE). A strategy profile $(\beta_1, \dots, \beta_n)$ is a BNE if, given the strategies of other players, no player can improve their expected utility by unilaterally deviating from their strategy. Formally, a strategy profile is a BNE if for each player $i$:
\[
\mathbb{E}[u_i(\beta_i(o_i, v_i), \beta_{-i}(o_{-i}, v_{-i}), v_i)] \geq \mathbb{E}[u_i(a_i, \beta_{-i}(o_{-i}, v_{-i}), v_i)] \quad \forall a_i \in A_i.
\]
This ensures that each player's strategy is optimal, given the strategies of others and their private information.

An \(\epsilon\)-Bayesian Nash Equilibrium (\(\epsilon\)-BNE) relaxes this condition slightly. A strategy profile is an \(\epsilon\)-BNE if no player can improve their expected utility by more than \(\epsilon \geq 0\) by deviating from their strategy. Formally, $(\beta_1, \dots, \beta_n)$ is an \(\epsilon\)-BNE if for each player $i$:
\[
\mathbb{E}[u_i(\beta_i(o_i, v_i), \beta_{-i}(o_{-i}, v_{-i}), v_i)] \geq \mathbb{E}[u_i(a_i, \beta_{-i}(o_{-i}, v_{-i}), v_i)] - \epsilon \quad \forall a_i \in A_i.
\]
This allows for small deviations in optimality, but still ensures that no player can significantly improve their outcome by deviating.

\subsection{Example: First-Price Auction with Two Bidders}

Consider a first-price sealed-bid auction with two bidders, each with a private valuation drawn from a uniform distribution on $[0,1]$. The bidders simultaneously submit bids, and the highest bidder wins and pays their bid. Each bidder’s objective is to maximize their expected utility, defined as the difference between their valuation and their bid if they win.

Let bidder $i$'s valuation $v_i \sim U[0,1]$. The utility for bidder $i$ is:
\[
u_i(b_i, b_j, v_i) = 
\begin{cases} 
v_i - b_i & \text{if } b_i > b_j, \\
0 & \text{if } b_i < b_j.
\end{cases}
\]
In the symmetric Bayesian Nash equilibrium, the bidding strategy is known to be $\beta(v) = \frac{v}{2}$. This strategy ensures that both bidders bid half of their valuation, which maximizes their expected utility, given the other bidder follows the same strategy. This illustrates the strategic decision-making in auctions within a Bayesian framework.

\section{Auctions in the Reinforcement Learning Framework}

Reinforcement learning (RL) provides a framework for agents to learn optimal strategies through interaction with an environment. We recast Bayesian games, specifically auctions, into the RL framework, allowing bidders to learn equilibrium strategies through repeated play. The goal is for each bidder to converge toward a Bayesian Nash equilibrium by applying RL algorithms.

In RL, an agent interacts with the environment over discrete time steps. At each time step $t$, the agent observes a state $s_t$, selects an action $a_t$ according to a policy $\pi(a_t | s_t)$, receives a reward $r_t$, and transitions to a new state $s_{t+1}$. The key components in the context of auctions are:

\subsection{States, Actions, and Rewards}

The \emph{state} of player $i$, $s_i \in S_i = V_i \times O_i$, includes their private valuation $v_i$ and any observations $o_i$. In sealed-bid auctions, $o_i$ may be empty; in dynamic auctions, it may include previous bids or signals.

The \emph{action} $a_i \in A_i$ is the bid chosen by player $i$. The action space can be continuous or discrete, depending on the auction design.

The \emph{reward} function $r_i(s_i, a_i)$ depends on the auction outcome, determined by the actions of all players. In a first-price auction, the reward for player $i$ is:
\[
r_i(s_i, a_i) = 
\begin{cases} 
v_i - a_i, & \text{if } a_i > \max_{j \neq i} a_j, \\
\frac{v_i - a_i}{k}, & \text{if } a_i = \max_{j \neq i} a_j \text{ and tied among } k \text{ bidders}, \\
0, & \text{otherwise}.
\end{cases}
\]

In multi-stage auctions, the state may include additional components like the round number or action history. While valuations $v_i$ are fixed in an auction, the state evolves as the auction progresses.

\subsection{Policy and Value Functions}

Each agent maintains a \emph{policy} $\pi_i(a_i | s_i)$, defining the probability distribution over actions based on the current state. The policy is parameterized (e.g., by $\theta_i$) and updated over time to maximize expected rewards. Initially, policies are stochastic to encourage exploration.

Agents estimate a \emph{value function} $V^{\pi_i}(s_i)$, representing the expected return when starting from state $s_i$ and following policy $\pi_i$. This helps assess state quality and reduces variance in policy updates.

\subsection{Learning Process}

The learning involves episodes, each corresponding to an independent auction. At each episode:

\begin{enumerate}
    \item \textbf{State Observation}: Agent $i$ observes $s_i = (v_i, o_i)$.
    \item \textbf{Action Selection}: Agent selects $a_i$ according to $\pi_i(a_i | s_i)$.
    \item \textbf{Auction Outcome}: Actions determine the outcome; agent receives reward $r_i$.
    \item \textbf{Experience Collection}: Agent collects $(s_i, a_i, r_i, s_i')$; in single-shot auctions, $s_i'$ may not exist.
    \item \textbf{Policy Update}: Agent updates policy parameters using collected experience.
\end{enumerate}

In multi-agent settings, each agent's optimal policy depends on others' policies. Convergence requires agents' policies to stabilize, which is challenging since the environment is non-stationary from any single agent's perspective. Therefore, 

\subsection{Example: First-Price Auction}

In a first-price auction with two bidders, each bidder's valuation $v_i$ is drawn from $[0,1]$. The state is $s_i = v_i$, and the action is the bid $a_i$. The reward is as defined earlier. Each bidder maintains a policy $\pi_i(a_i | s_i; \theta_i)$. Through repeated auctions, bidders collect experiences $(s_i, a_i, r_i)$ and update $\theta_i$ using policy gradients. Initially random policies ensure exploration. Over time, policies converge toward the equilibrium strategy $\beta(v_i) = \frac{1}{2} v_i$ for two bidders.

\section{Algorithm}

\subsection{Policy Gradient Methods}

Policy gradient methods focus on directly optimizing the policy function, denoted as $\pi_\theta(a | s)$, with the objective of maximizing the expected cumulative return. The policy $\pi_\theta(a | s)$ represents the probability of taking action $a$ given state $s$, parameterized by the parameter vector $\theta$. The expected cumulative return is defined as:

\[
J(\theta) = \mathbb{E}_{\pi_\theta} \left[ \sum_{t=0}^{T} r_t \right],
\]

where $r_t$ is the reward received at time step $t$, and $T$ is the time horizon of an episode. To optimize this objective, the gradient of the expected return with respect to the policy parameters $\theta$ is computed as:

\[
\nabla_\theta J(\theta) = \mathbb{E}_{\pi_\theta} \left[ \sum_{t=0}^{T} \nabla_\theta \log \pi_\theta(a_t | s_t) A_t \right],
\]

where $A_t = A^{\pi_\theta}(s_t, a_t)$ is the advantage function. The advantage function measures the relative quality of action $a_t$ in state $s_t$ compared to the average performance of the policy in that state, effectively capturing how much better or worse an action is compared to others available in the same state.

Neural networks are typically employed to parameterize the policy $\pi_\theta(a | s)$. For environments with continuous action spaces, the policy outputs the parameters of a Gaussian distribution:

\[
\pi_\theta(a | s) = \mathcal{N}\left( \mu_\theta(s), \sigma_\theta(s) \right),
\]

where $\mu_\theta(s)$ and $\sigma_\theta(s)$ represent the mean and standard deviation of the distribution, respectively. In contrast, for discrete action spaces, the policy utilizes a softmax function to produce a probability distribution over the possible actions:

\[
\pi_\theta(a | s) = \frac{\exp(f_\theta(s, a))}{\sum_{a'} \exp(f_\theta(s, a'))}.
\]

\subsection{Proximal Policy Optimization (PPO)}

Proximal Policy Optimization (PPO) enhances the stability and reliability of policy gradient methods by constraining the extent to which the policy can change during each update. This is achieved through a clipped objective function that prevents large, potentially detrimental updates to the policy parameters. The core objective function of PPO is defined as:

\[
L^{\text{CLIP}}(\theta) = \mathbb{E}_t \left[ \min \left( r_t(\theta) A_t, \ \text{clip}\left( r_t(\theta), 1 - \epsilon, 1 + \epsilon \right) A_t \right) \right],
\]

where $r_t(\theta)$ is the probability ratio between the new policy and the old policy:

\[
r_t(\theta) = \frac{\pi_\theta(a_t | s_t)}{\pi_{\theta_{\text{old}}}(a_t | s_t)},
\]

and $\epsilon$ is a small hyperparameter that defines the permissible range for $r_t(\theta)$. The clipping function ensures that the ratio $r_t(\theta)$ remains within the interval $[1 - \epsilon, 1 + \epsilon]$. This mechanism effectively limits the policy update to a region where the new policy does not deviate excessively from the old policy, thereby maintaining training stability.

The clipped objective can be elaborated as follows:

\[
L^{\text{CLIP}}(\theta) = \mathbb{E}_t \left[ 
\begin{cases}
r_t(\theta) A_t, & \text{if } |r_t(\theta) - 1| \leq \epsilon \\
(1 + \epsilon) A_t, & r_t(\theta) > 1 + \epsilon, \ A_t \geq 0 \\
(1 - \epsilon) A_t, & r_t(\theta) < 1 - \epsilon, \ A_t \leq 0 \\
r_t(\theta) A_t, & \text{otherwise}
\end{cases}
\right].
\]

This formulation ensures a balance between exploration and exploitation by preventing the policy from making overly aggressive updates that could destabilize the learning process.

\subsection{Other Components}

Beyond the clipped objective, the PPO algorithm integrates several critical components to facilitate effective learning. The policy parameters, denoted by $\theta$, are updated by maximizing the clipped objective function in conjunction with an entropy regularization term, $\mathcal{H}(\pi_\theta(\cdot | s_t))$. The entropy term encourages exploration by penalizing policies that become too deterministic, thereby promoting a more diverse set of actions. The influence of the entropy term is controlled by the coefficient $\beta$, which determines the weight given to entropy regularization relative to the primary objective.

For each training iteration, a set of trajectories $\mathcal{D}_k = \{(s_t, a_t, r_t)\}$ is collected by interacting with the environment using the current policy $\pi_{\theta_k}$. From these trajectories, advantage estimates $\hat{A}_t$ and return estimates $\hat{R}_t$ are computed. The advantage estimates guide the policy updates by indicating the relative value of actions, while the return estimates are utilized to update the value function $V_\phi(s_t)$, parameterized by $\phi$. The value function is trained to minimize the value function loss $L^{\text{VF}}(\phi)$, defined as the mean squared error between the predicted values and the estimated returns:

\[
L^{\text{VF}}(\phi) = \frac{1}{|\text{mini-batch}|} \sum_t \left( V_\phi(s_t) - \hat{R}_t \right)^2.
\]

Both the policy and value function updates are performed using mini-batch stochastic gradient descent, iterating over subsets of the collected trajectories. This approach ensures efficient and stable learning by leveraging batch processing and gradient-based optimization. The value function aids in reducing the variance of the policy gradient estimates, while the advantage estimates refine the policy's action selection based on the relative performance of actions in specific states.

\subsection{Pseudocode}

To consolidate the aforementioned components, the PPO algorithm is summarized in the following pseudocode. 

\begin{algorithm}[H]
\caption{Proximal Policy Optimization (PPO)}
\label{alg:ppo}
\textbf{Initialize}: Policy parameters $\theta_0$, Value function parameters $\phi_0$, Clipping parameter $\epsilon$, Policy learning rate $\alpha_\theta$, Value function learning rate $\alpha_\phi$, Entropy regularization coefficient $\beta$, Total number of iterations $K$ \\

\For{$k = 1$ to $K$}{
    \textbf{Collect Trajectories}: 
    \[
    \mathcal{D}_k = \{(s_t, a_t, r_t)\}_{t=0}^{T}
    \]
    using the current policy $\pi_{\theta_k}$.

    \textbf{Compute Estimates}: 
    \[
    \hat{A}_t = A^{\pi_{\theta_k}}(s_t, a_t), \quad \hat{R}_t = \sum_{t'=t}^{T} \gamma^{t'-t} r_{t'}
    \]
    where $\gamma$ is the discount factor.

    \For{each mini-batch within $\mathcal{D}_k$}{
        \textbf{Probability Ratio}: 
        \[
        r_t(\theta) = \frac{\pi_\theta(a_t | s_t)}{\pi_{\theta_k}(a_t | s_t)}.
        \]

        \textbf{Clipped Objective}: 
        \[
        L^{\text{CLIP}}(\theta) = \mathbb{E}_t \left[ \min \left( r_t(\theta) \hat{A}_t, \ \text{clip}\left( r_t(\theta), 1 - \epsilon, 1 + \epsilon \right) \hat{A}_t \right) \right].
        \]

        \textbf{Policy Update}: 
        \[
        \theta \leftarrow \theta + \alpha_\theta \nabla_\theta \left( L^{\text{CLIP}}(\theta) - \beta \mathcal{H} \left( \pi_\theta(\cdot | s_t) \right) \right),
        \]
        where $\mathcal{H}$ denotes the entropy of the policy to encourage exploration.

        \textbf{Value Function Update}: 
        \[
        \phi \leftarrow \phi - \alpha_\phi \nabla_\phi L^{\text{VF}}(\phi),
        \]
        where 
        \[
        L^{\text{VF}}(\phi) = \frac{1}{|\text{mini-batch}|} \sum_t \left( V_\phi(s_t) - \hat{R}_t \right)^2
        \]
        represents the value function loss, aiming to minimize the discrepancy between the predicted values and the actual returns.
    }
}
\end{algorithm}

\section{Experiments}

\subsection{First Price Auction}

In a first price auction with two bidders and independent private valuations drawn from a uniform distribution \( v \sim \text{UNIF}(0,1) \), the symmetric Bayesian Nash Equilibrium (BNE) bidding strategy is \( b^*(v) = \frac{v}{2} \).

\begin{figure}[H]
    \centering
    \includegraphics[width=0.5\textwidth]{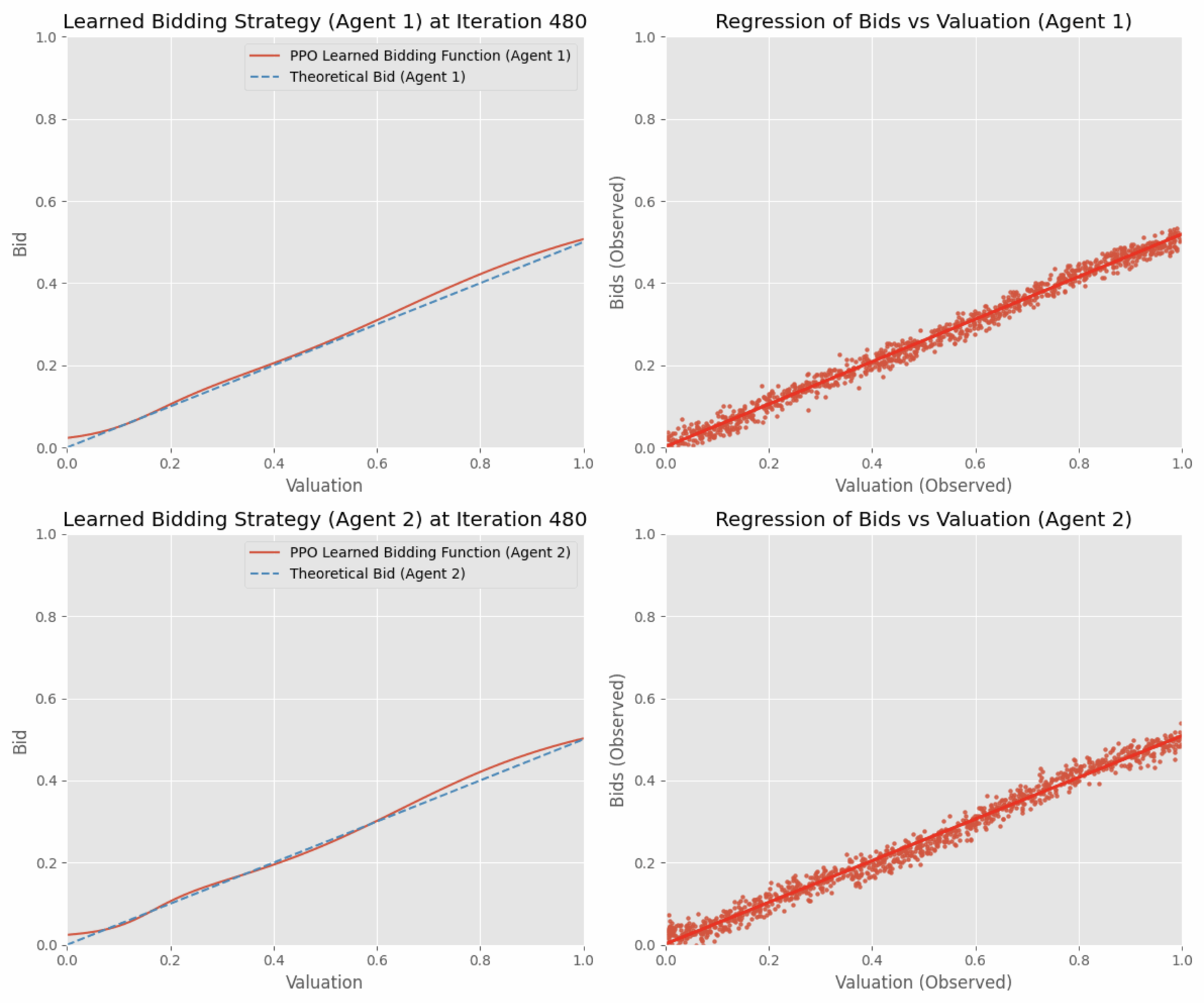}
    \label{fig:fpa}
\end{figure}

\subsection{First Price Auction with Power Distribution}

Here, bidders' valuations are drawn from a power distribution \( f(v) = v^{1/2} \). The symmetric BNE bidding strategy is \( b^*(v) = \frac{v}{3} \).

\begin{figure}[H]
    \centering
    \includegraphics[width=0.5\textwidth]{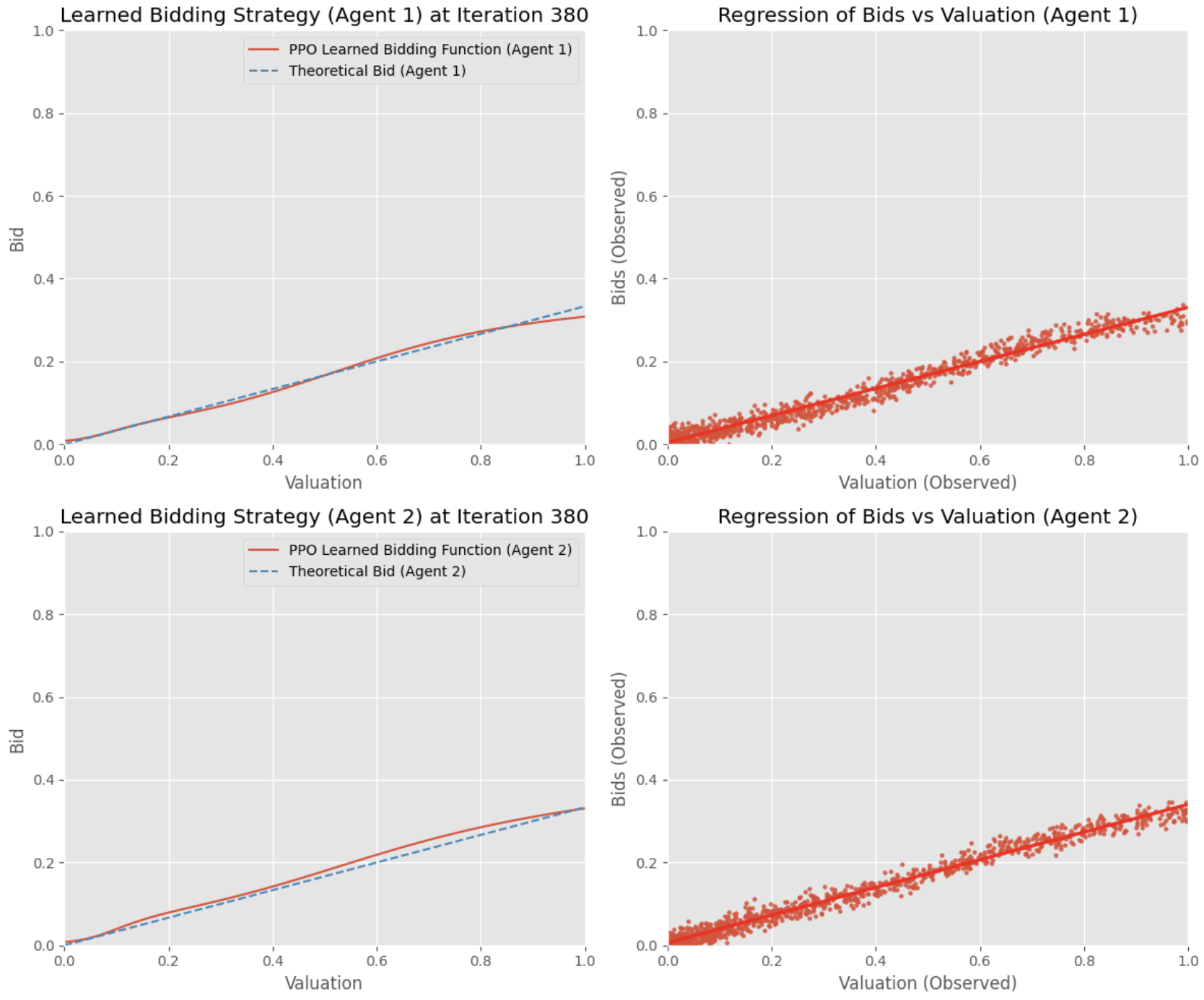}
    \label{fig:fpa_power}
\end{figure}

\subsection{First Price Auction with Risk Aversion}

In this setting, bidders are risk-averse, and the utility of the winner is given by \( u_i = \sqrt{v_i - b_i} \). The symmetric BNE bidding strategy becomes: $b^*(v) = 2v/3.$

\begin{figure}[H]
    \centering
    \includegraphics[width=0.5\textwidth]{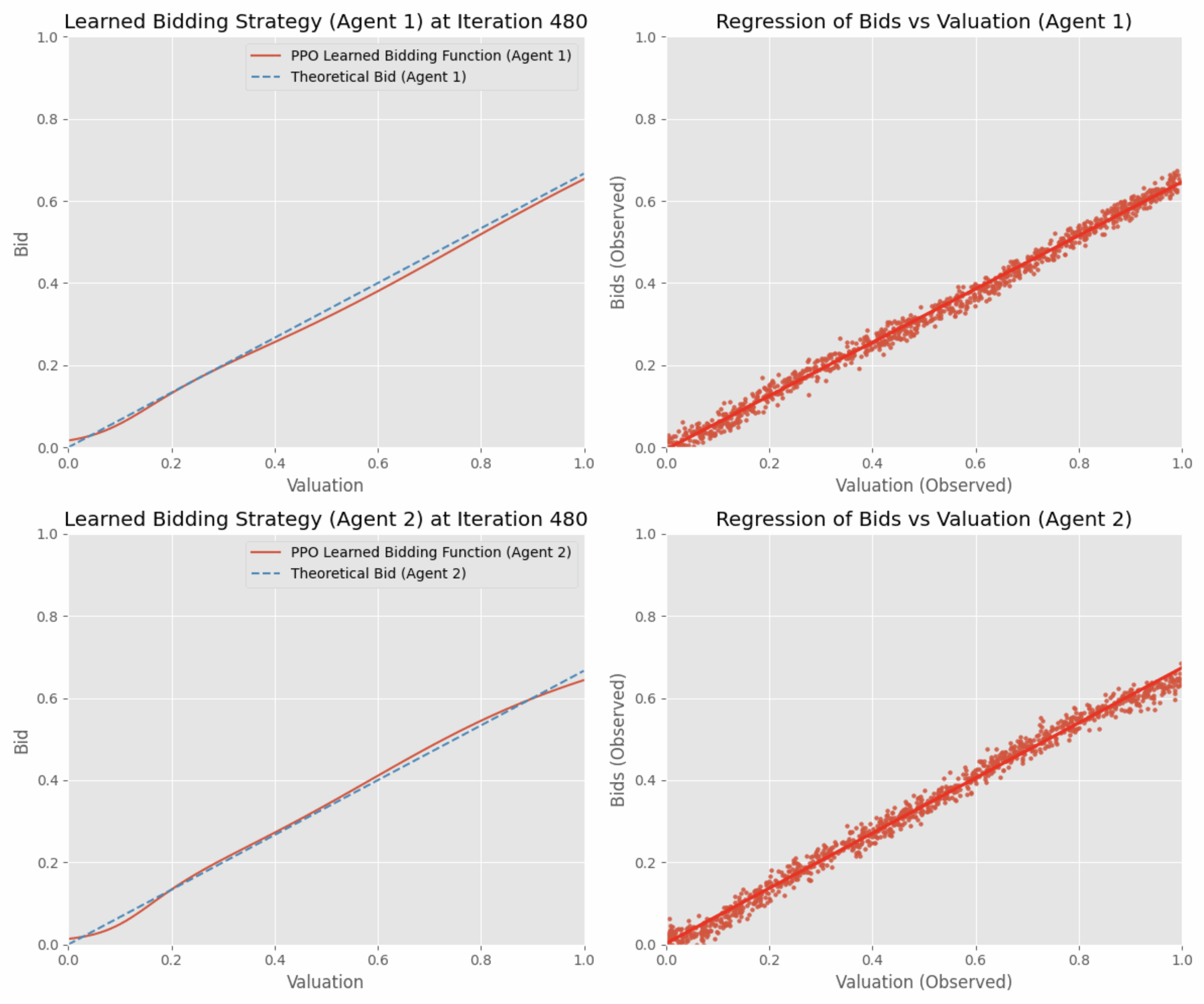}
    \label{fig:fpa_ra}
\end{figure}

\subsection{First Price Auction with Asymmetric Bidders}

Bidders have asymmetric valuation distributions: Agent 1's valuation \( v_1 \sim \text{UNIF}(0,1.33) \), and Agent 2's valuation \( v_2 \sim \text{UNIF}(0,0.8) \). The asymmetric BNE is complex and does not have a simple closed-form expression.

\begin{figure}[H]
    \centering
    \includegraphics[width=0.65\textwidth]{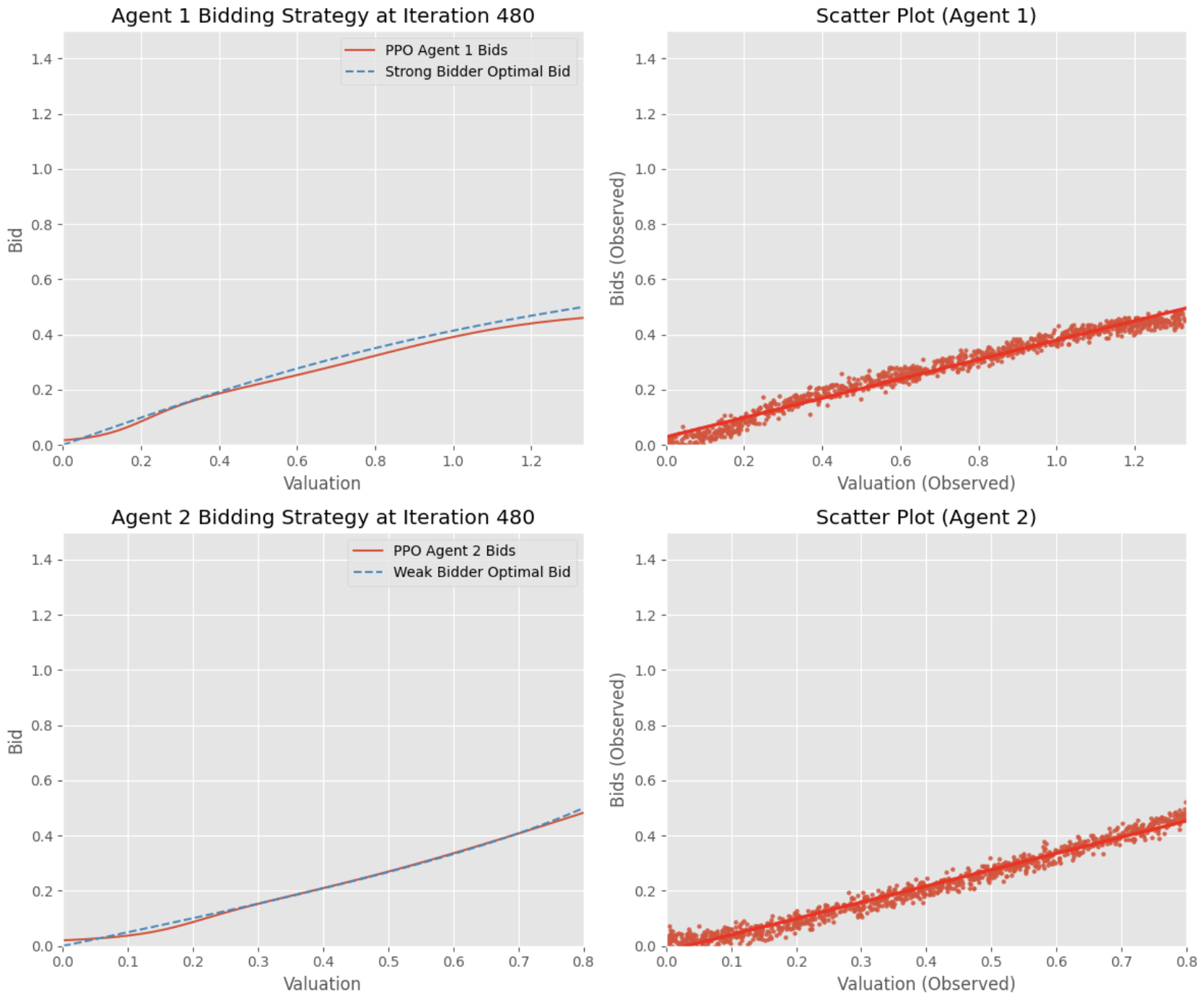}
    \label{fig:fpa_asym}
\end{figure}

\subsection{First Price Auction with Reserve Price}

A reserve price \( r = 0.25 \) is introduced below which bidders automatically lose. Valuations are drawn from \( v \sim \text{UNIF}(0,1) \).

\begin{figure}[H]
    \centering
    \includegraphics[width=0.5\textwidth]{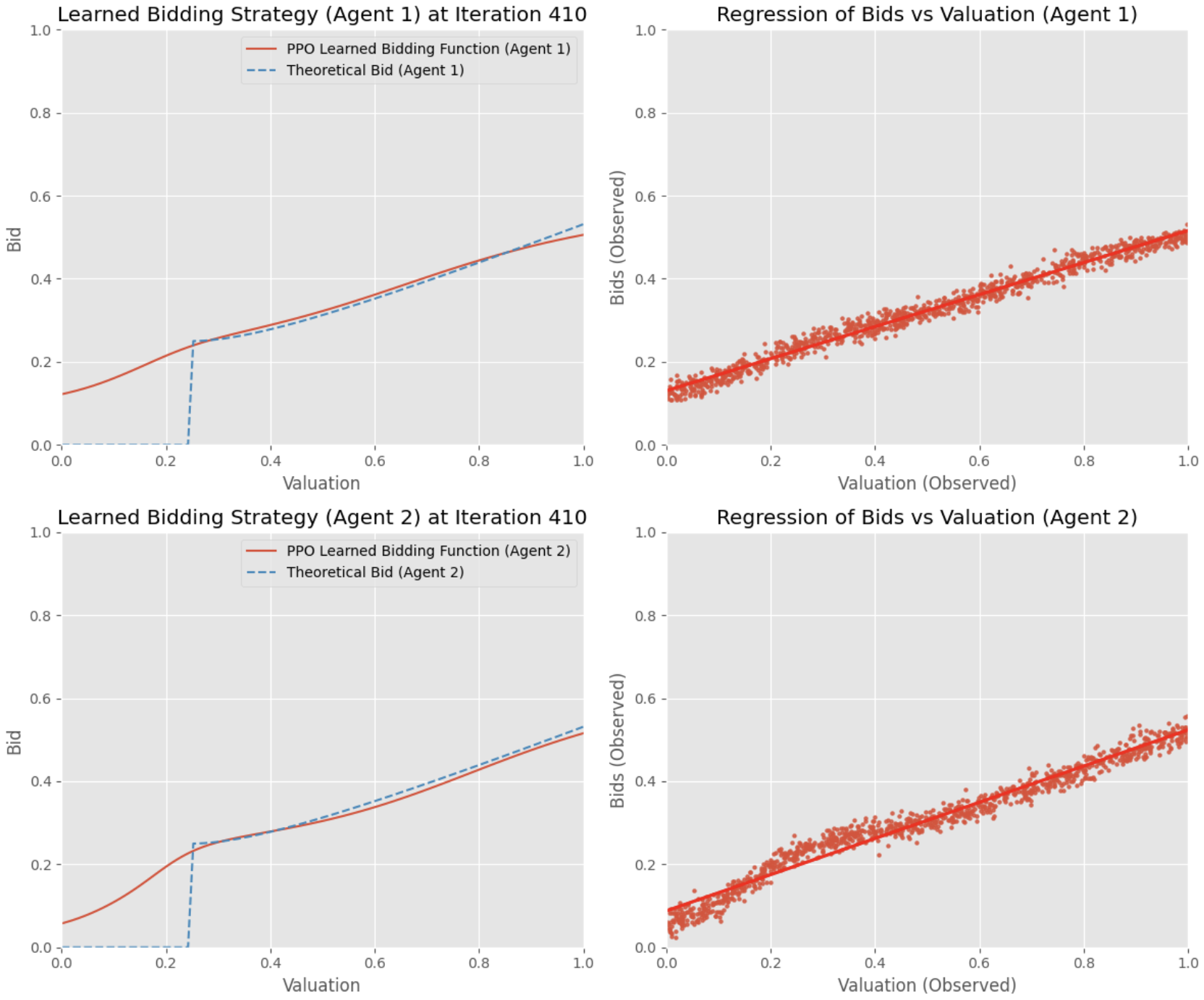}
    \label{fig:fpa_res}
\end{figure}

\subsection{Second Price Auction}

In a second price auction with two bidders and valuations \( v \sim \text{UNIF}(0,1) \), the symmetric BNE bidding strategy is truthful bidding: $b^*(v) = v.$
\begin{figure}[H]
    \centering
    \includegraphics[width=0.6\textwidth]{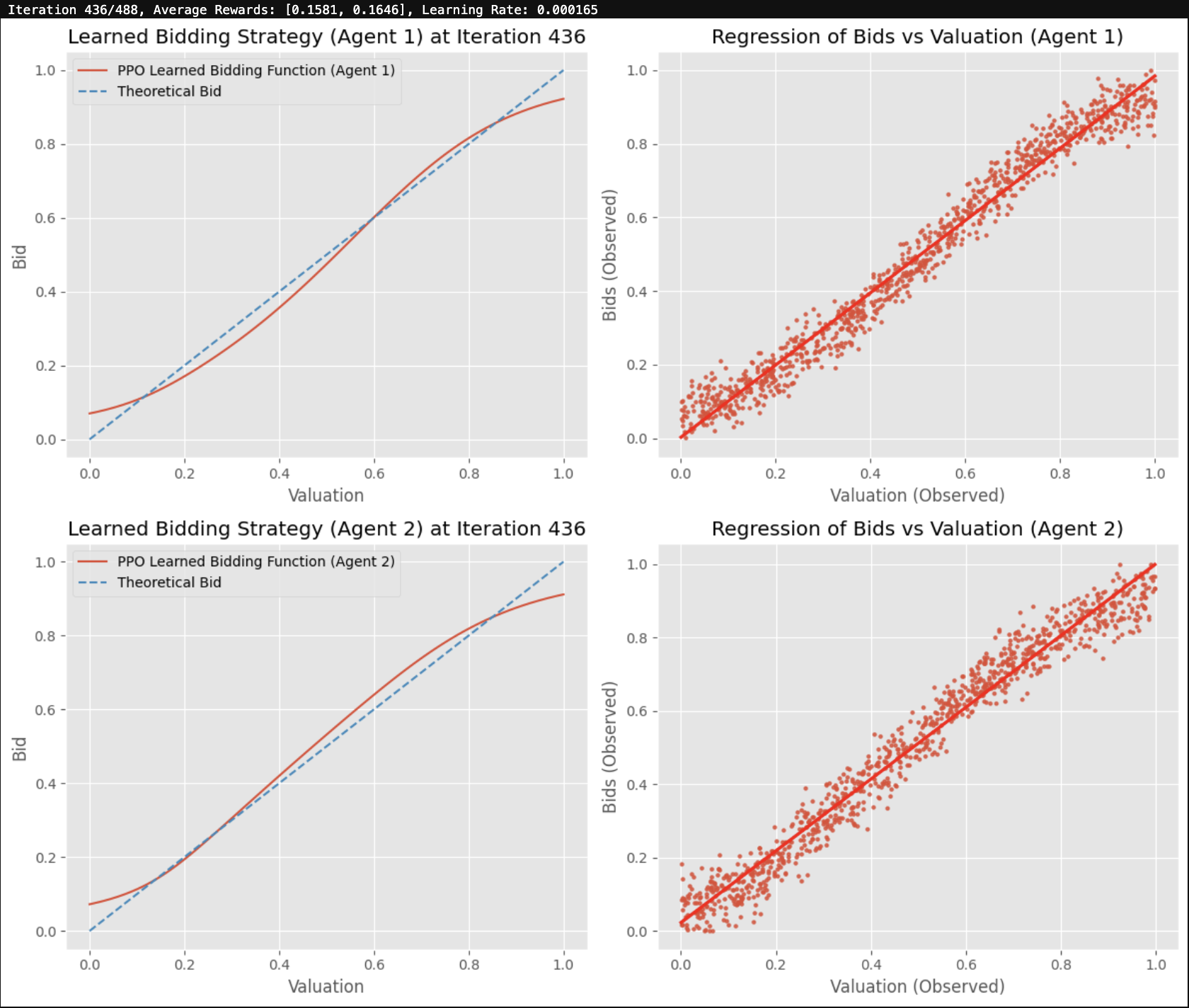}
    \label{fig:spa}
\end{figure}

\subsection{All-Pay Auction}

In an all-pay auction with two bidders and valuations \( v \sim \text{UNIF}(0,1) \), the symmetric BNE bidding strategy is: $b^*(v) = v^2/2$

\begin{figure}[H]
    \centering
    \includegraphics[width=0.6\textwidth]{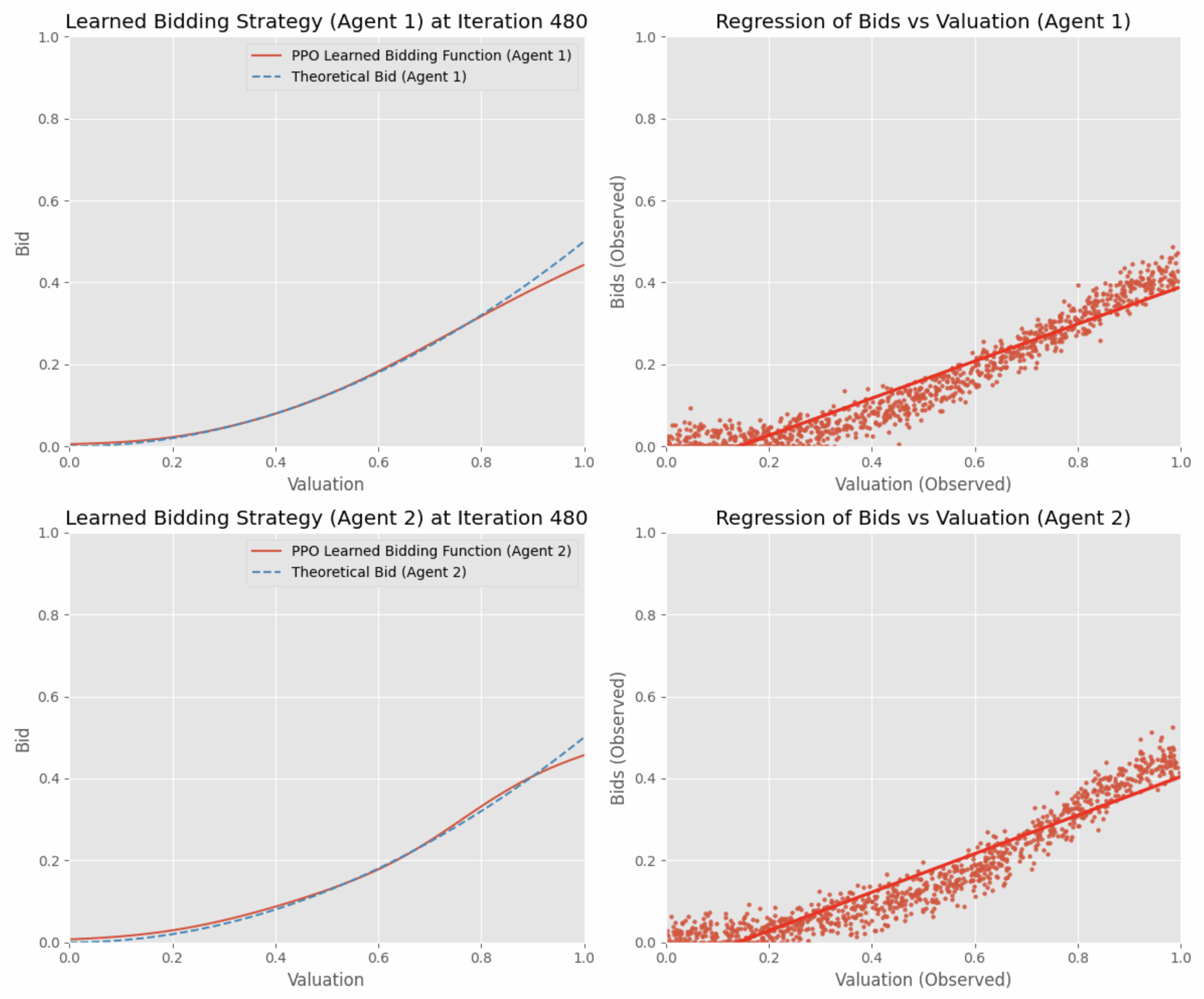}
    \label{fig:apa}
\end{figure}

\subsection{Third-Price Auction}

With three bidders and valuations \( v \sim \text{UNIF}(0,1) \), the symmetric BNE bidding strategy in a third-price auction is: $b^*(v) = 2v$.

\begin{figure}[H]
    \centering
    \includegraphics[width=0.5\textwidth]{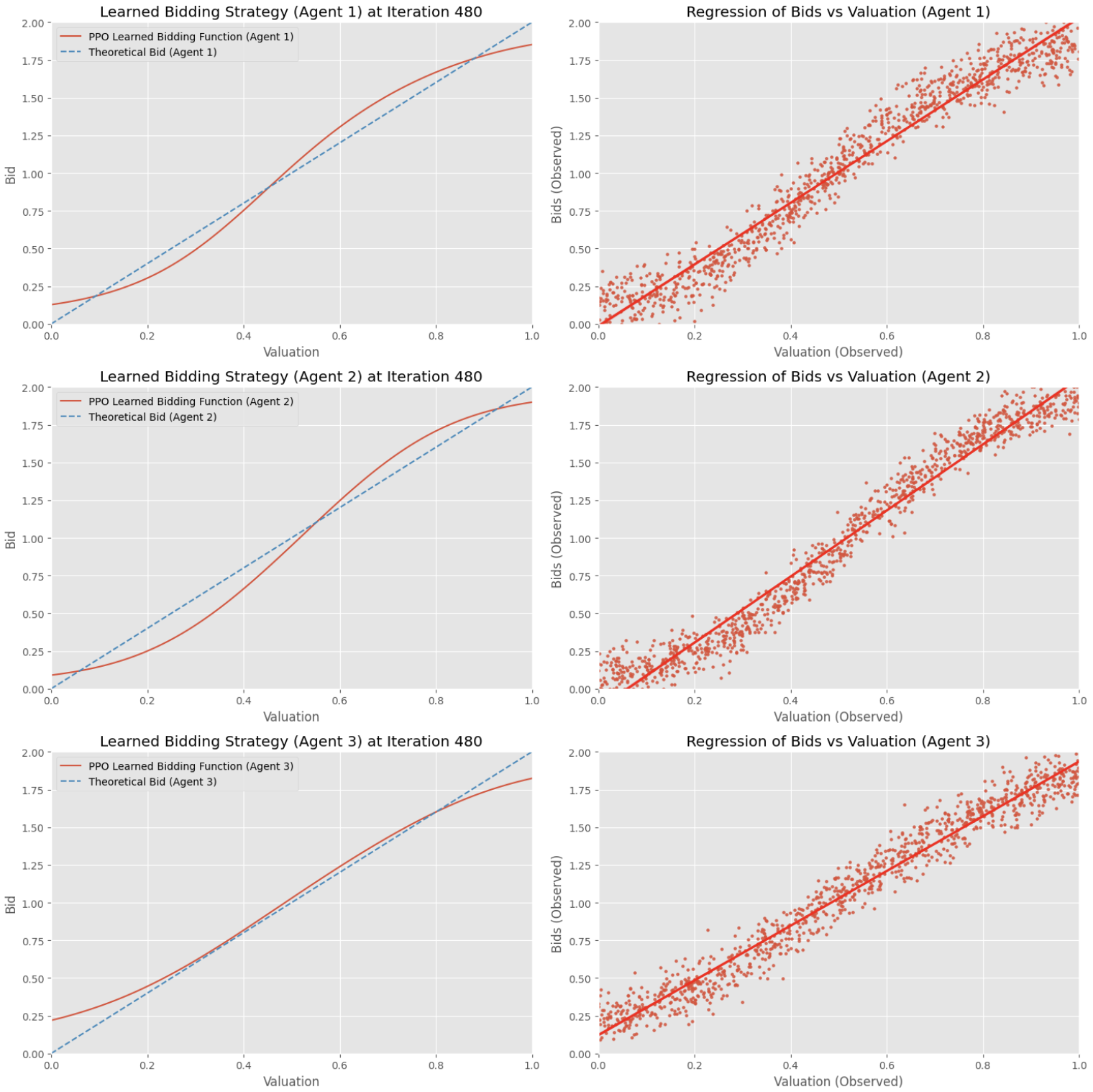}
    \label{fig:tpa}
\end{figure}

\subsection{First Price Auction with Common Values}

Bidders receive signals \( x_i = k_i + t \) where \( k_i, t \sim \text{UNIF}(0,1) \), and the common value is \( v = \frac{x_1 + x_2}{2} \). The symmetric BNE bidding strategy is:$b^*(x) = 2x/3$.

\begin{figure}[H]
    \centering
    \includegraphics[width=0.5\textwidth]{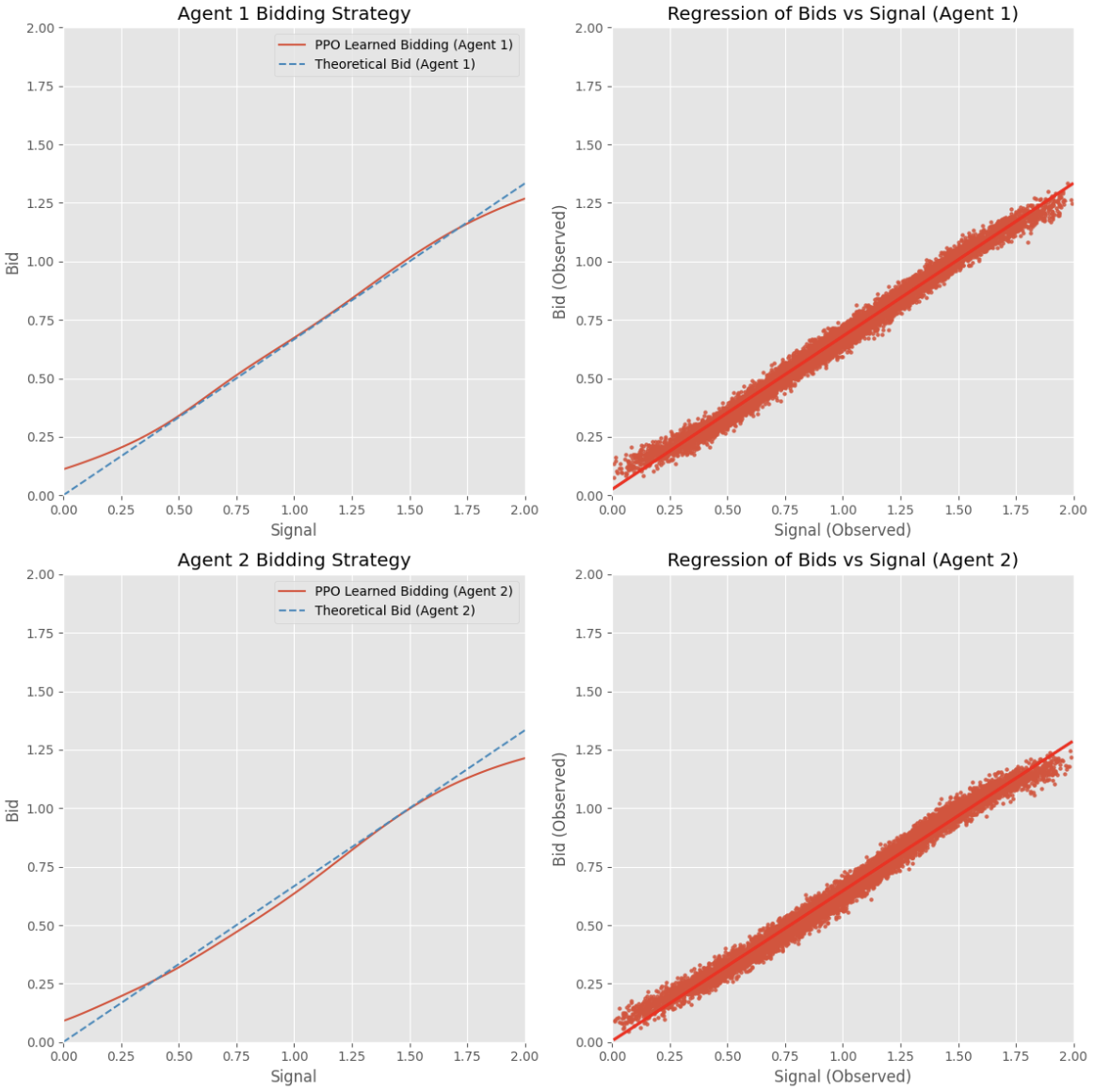}
    \label{fig:fpa_cv}
\end{figure}

\subsection{Second Price Auction with Common Values}

With three bidders, valuations \( v \sim \text{UNIF}(0,1) \), and independent signals \( x \sim \text{UNIF}(0,2v) \), the symmetric BNE bidding strategy is: $b^*(x) = 2x/(2+x)$.
\begin{figure}[H]
    \centering
    \includegraphics[width=0.5\textwidth]{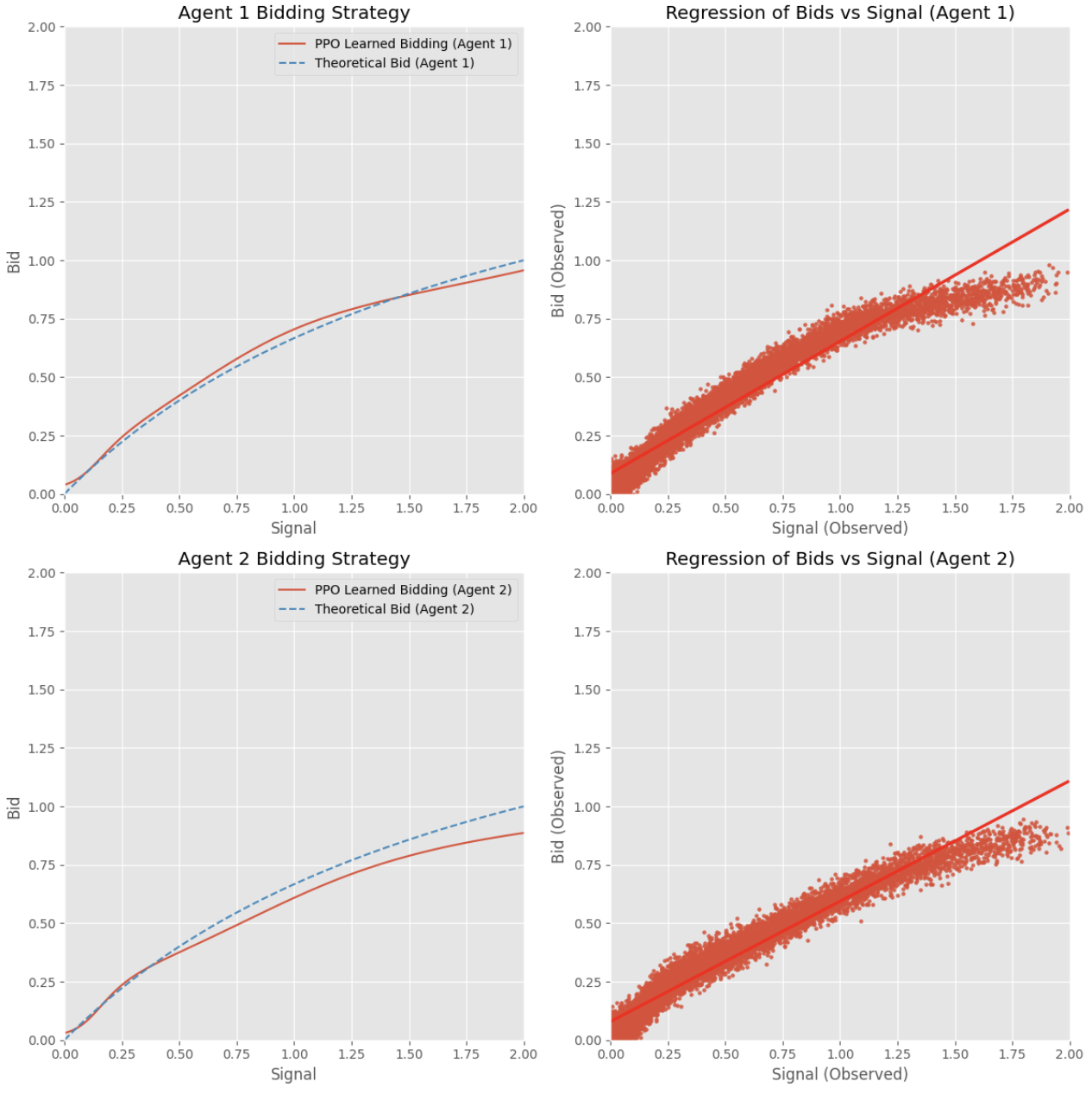}
    \label{fig:spa_cv}
\end{figure}

\subsection{Korean Auction}

In the Korean auction, two bidders have private valuations \( v \sim \text{UNIF}(0,1) \). They receive a signal \( x = 1 \) if the bidder holds the highest bid in round 0; otherwise, \( x = 0 \). The auction proceeds in rounds indexed by \( t = 0,1 \).

\begin{figure}[H]
    \centering
    \includegraphics[width=0.5\textwidth]{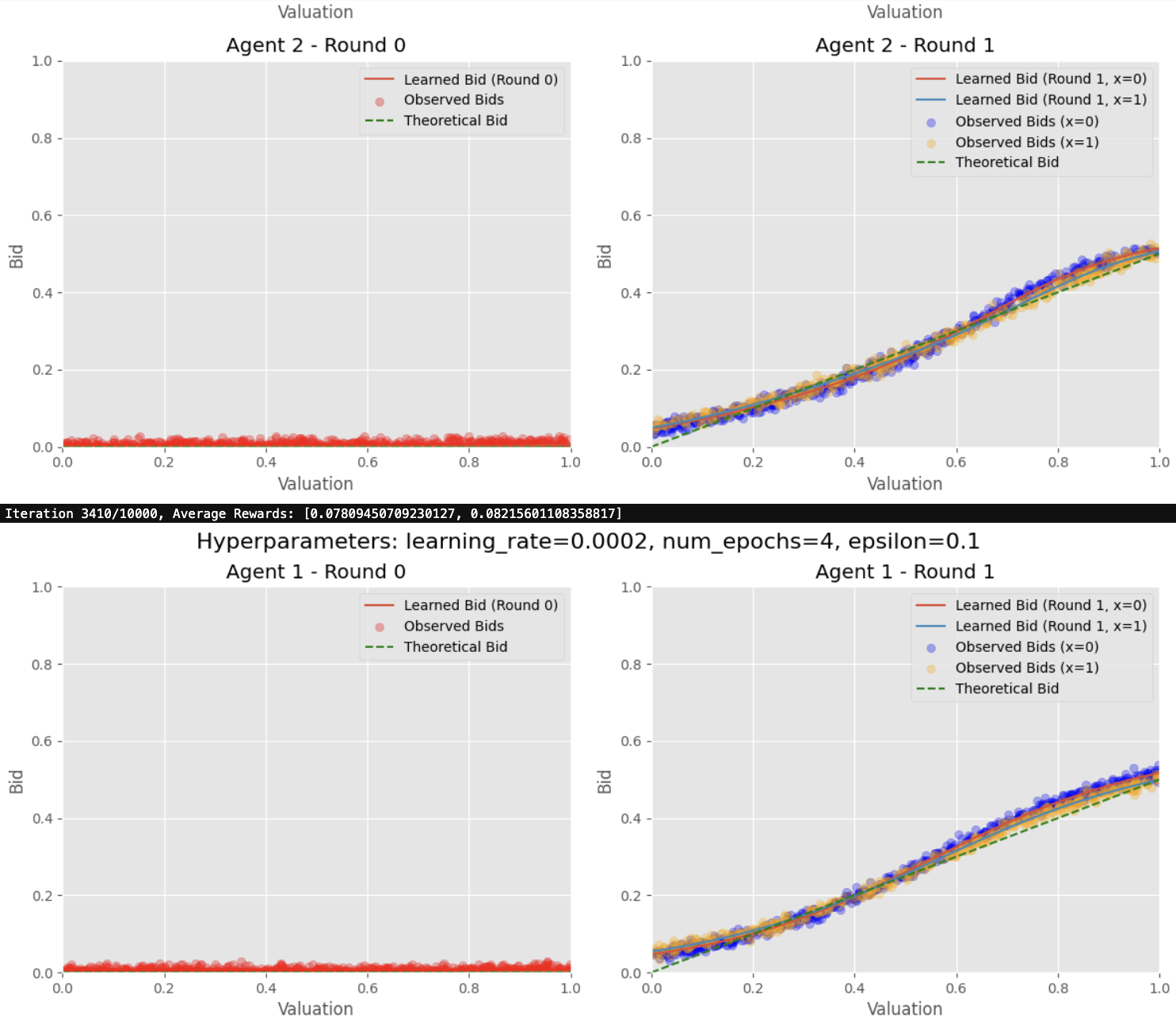}
    \label{fig:korean}
\end{figure}

\section{Conclusion}

In this study, we presented an RL-based approach to approximate equilibria in auctions, offering a framework for bidders to learn optimal strategies through repeated interactions. We validated this approach against theoretical benchmarks, showing its effectiveness in simple auction settings such as first-price auctions. 

There are several directions for future work. First, a robust hyperparameter configuration needs to be established, ensuring the approach works consistently across different types of auctions. Addressing the issue of policy collapse as standard deviation declines is critical, possibly requiring more sophisticated methods like robust learning rate annealing or second-order methods such as Trust Regions. Incorporating model-based insights could improve sample efficiency and accelerate convergence. Lastly, extending the framework to handle high-dimensional action spaces, such as those in simultaneous auctions, will be an essential step in scaling the approach to more complex auction environments.

\section{Appendix: Policy Gradient Algorithms}

We aim to optimize a stochastic policy $\pi_\theta(a|s)$ parameterized by $\theta$ to maximize the expected return:
\[
J(\theta) = \mathbb{E}_{\tau \sim P(\tau|\theta)} [R(\tau)],
\]
where:
\begin{itemize}
    \item $\tau = (s_0, a_0, s_1, a_1, \dots, s_T, a_T)$ is a trajectory,
    \item $R(\tau) = \sum_{t=0}^{T} r(s_t, a_t)$ is the cumulative reward,
    \item $P(\tau|\theta)$ is the probability of $\tau$ under policy $\pi_\theta$.
\end{itemize}

Our goal is to derive an expression for the gradient $\nabla_\theta J(\theta)$ that can be estimated from samples and used in gradient ascent algorithms.

\subsection{Probability of a Trajectory}

The probability of a trajectory $\tau$ under policy $\pi_\theta$ is:
\[
P(\tau|\theta) = P(s_0) \prod_{t=0}^{T-1} \pi_\theta(a_t|s_t) P(s_{t+1}|s_t, a_t),
\]
where:
\begin{itemize}
    \item $P(s_0)$ is the initial state distribution,
    \item $P(s_{t+1}|s_t, a_t)$ is the environment transition probability, independent of $\theta$.
\end{itemize}

\subsection{The Log-Derivative Trick}

The log-derivative trick is based on the identity:
\[
\nabla_\theta f(\theta) = f(\theta) \nabla_\theta \log f(\theta),
\]
for any differentiable function $f(\theta) > 0$.

\subsection{Gradient of Expected Return}

We compute the gradient of $J(\theta)$:
\[
\nabla_\theta J(\theta) = \nabla_\theta \mathbb{E}_{\tau \sim P(\tau|\theta)} [R(\tau)] = \nabla_\theta \int P(\tau|\theta) R(\tau) \, d\tau.
\]

Differentiating under the integral sign:
\[
\nabla_\theta J(\theta) = \int \nabla_\theta P(\tau|\theta) R(\tau) \, d\tau.
\]

\subsection{Applying the Log-Derivative Trick}

Using the log-derivative trick:
\[
\nabla_\theta P(\tau|\theta) = P(\tau|\theta) \nabla_\theta \log P(\tau|\theta).
\]

Thus:
\[
\nabla_\theta J(\theta) = \int P(\tau|\theta) \nabla_\theta \log P(\tau|\theta) R(\tau) \, d\tau = \mathbb{E}_{\tau \sim P(\tau|\theta)} [R(\tau) \nabla_\theta \log P(\tau|\theta)].
\]

\subsection{Computing $\nabla_\theta \log P(\tau|\theta)$}

Since the environment dynamics do not depend on $\theta$:
\begin{align*}
\log P(\tau|\theta) &= \log P(s_0) + \sum_{t=0}^{T-1} \left[ \log \pi_\theta(a_t|s_t) + \log P(s_{t+1}|s_t, a_t) \right], \\
\nabla_\theta \log P(\tau|\theta) &= \sum_{t=0}^{T-1} \nabla_\theta \log \pi_\theta(a_t|s_t).
\end{align*}

\subsection{Policy Gradients}

The simplest policy gradient can be obtained by substituting back:
\[
\nabla_\theta J(\theta) = \mathbb{E}_{\tau \sim P(\tau|\theta)} \left[ R(\tau) \sum_{t=0}^{T-1} \nabla_\theta \log \pi_\theta(a_t|s_t) \right].
\]

\subsection{Expected Grad-Log-Prob Lemma}

For any parameterized probability distribution $p_\theta(x)$,
\[
\mathbb{E}_{x \sim p_\theta} [\nabla_\theta \log p_\theta(x)] = 0.
\]

\textbf{Proof}:

Starting from the normalization condition:
\[
\int p_\theta(x) \, dx = 1.
\]
Differentiating both sides with respect to $\theta$:
\[
\int \nabla_\theta p_\theta(x) \, dx = 0.
\]
Using the log-derivative trick:
\[
\int p_\theta(x) \nabla_\theta \log p_\theta(x) \, dx = 0.
\]
Therefore:
\[
\mathbb{E}_{x \sim p_\theta} [\nabla_\theta \log p_\theta(x)] = 0.
\]

This lemma allows us to adjust the policy gradient by adding terms that integrate to zero, aiding in variance reduction.

\subsection{Reward-to-Go Policy Gradient}

In the simplest policy gradient, actions are reinforced based on the total return $R(\tau)$, which includes rewards from before and after the action. However, actions cannot affect past rewards. So we can replace $R(\tau)$ with the reward-to-go from time $t$:

\[
R_t = \sum_{t'=t}^{T} r(s_{t'}, a_{t'}).
\]
This gives:
\[
\nabla_\theta J(\theta) = \mathbb{E}_{\tau \sim P(\tau|\theta)} \left[ \sum_{t=0}^{T-1} R_t \nabla_\theta \log \pi_\theta(a_t|s_t) \right].
\]

Subtracting the expected cumulative reward before time $t$ (which does not depend on $a_t$) does not introduce bias, due to the Expected Grad-Log-Prob Lemma.

\subsection{Variance Reduction via Baselines}

We can subtract a baseline $b(s_t)$ from $R_t$ to reduce variance:
\[
\nabla_\theta J(\theta) = \mathbb{E}_{\tau \sim P(\tau|\theta)} \left[ \sum_{t=0}^{T-1} \left( R_t - b(s_t) \right) \nabla_\theta \log \pi_\theta(a_t|s_t) \right].
\]

Baselines do not introduce bias. We can see this from the Expected Grad-Log-Prob Lemma:
\[
\mathbb{E}_{\tau \sim P(\tau|\theta)} \left[ \sum_{t=0}^{T-1} b(s_t) \nabla_\theta \log \pi_\theta(a_t|s_t) \right] = 0,
\]
since $b(s_t)$ does not depend on $a_t$. How do we choose the baseline? A common choice is the state-value function:
\[
b(s_t) = V^\pi(s_t) = \mathbb{E}_{\pi} [R_t | s_t].
\]
Subtracting $V^\pi(s_t)$ reduces variance by centering the returns.

\subsection{Other Baselines}

We can express the policy gradient using the action-value function $Q^\pi(s_t, a_t)$:
\[
\nabla_\theta J(\theta) = \mathbb{E}_{\pi} \left[ \sum_{t=0}^{T-1} Q^\pi(s_t, a_t) \nabla_\theta \log \pi_\theta(a_t|s_t) \right].
\]

Or, we can use the advantage function:
\[
A^\pi(s_t, a_t) = Q^\pi(s_t, a_t) - V^\pi(s_t),
\]
the policy gradient becomes:
\[
\nabla_\theta J(\theta) = \mathbb{E}_{\pi} \left[ \sum_{t=0}^{T-1} A^\pi(s_t, a_t) \nabla_\theta \log \pi_\theta(a_t|s_t) \right].
\]

\subsection{Value Function Approximation}

In practice, the value function $V^\pi(s_t)$ cannot be computed exactly, so it is approximated using a neural network, $V_\phi(s_t)$, which is updated concurrently with the policy. The simplest method for learning $V_\phi$ is to minimize the mean-squared-error (MSE) between the predicted value and the empirical return:
\[
\phi_k = \arg \min_{\phi} \mathbb{E}_{s_t, \hat{R}_t \sim \pi_k} \left[ \left( V_\phi(s_t) - \hat{R}_t \right)^2 \right],
\]
where $\pi_k$ is the policy at epoch $k$. This is done using one or more steps of gradient descent starting from the previous parameters $\phi_{k-1}$.

\subsection{Vanilla Policy Gradient}

The simplest reinforcement learning algorithms that uses policy gradients can be expressed in the following way. The policy gradient is computed as:
\[
\nabla_\theta J(\pi_\theta) = \mathbb{E}_{\tau \sim \pi_\theta} \left[ \sum_{t=0}^{T} \nabla_\theta \log \pi_\theta(a_t|s_t) A^{\pi_\theta}(s_t, a_t) \right],
\]

The algorithm updates the policy parameters via stochastic gradient ascent:
\[
\theta_{k+1} = \theta_k + \alpha \nabla_\theta J(\pi_\theta),
\]
where $\alpha$ is the learning rate.

\subsection{Reusing Old Data using Importance Sampling}

We can reverse this gradient expression and define the corresponding objective as:
\[
L^{PG}(\theta) = \mathbb{E}_t \left[ \log \pi_\theta(a_t | s_t) \hat{A}_t \right].
\]

Using importance sampling using older policies \( \pi_{\text{old}} \), the same objective is:
\[
L^{IS}_{\theta_{\text{old}}}(\theta) = \mathbb{E}_t \left[ \frac{\pi_\theta(a_t | s_t)}{\pi_{\theta_{\text{old}}}(a_t | s_t)} \hat{A}_t \right].
\]
This allows us to reuse data collected under the old policy \( \pi_{\text{old}} \).

Note that the gradients of both objective functions are the same, which means they lead to the same gradient ascent:
\[
\nabla_\theta \log \pi_\theta(a_t | s_t) \bigg|_{\theta_{\text{old}}} = \nabla_\theta \left( \frac{\pi_\theta(a_t | s_t)}{\pi_{\text{old}}(a_t | s_t)} \right) \bigg|_{\theta_{\text{old}}}.
\]

\subsection{Trust Region Policy Optimization (TRPO)}

Vanilla policy gradients can suffer from instability due to large updates to the policy parameters. TRPO addresses this by enforcing a trust region constraint, measured by the Kullback-Leibler (KL) divergence, on the policy update.

The optimization problem for TRPO is:
\[
\max_{\theta'} \mathbb{E}_{\tau \sim \pi_\theta} \left[ \frac{\pi_{\theta'}(a_t | s_t)}{\pi_\theta(a_t | s_t)} A^{\pi_\theta}(s_t, a_t) \right]
\]
subject to:
\[
\mathbb{E}_{s \sim \pi_\theta} \left[ D_{\text{KL}} \left( \pi_\theta(\cdot|s) \| \pi_{\theta'}(\cdot|s) \right) \right] \leq \delta,
\]
where $D_{\text{KL}}$ is the KL divergence, and $\delta$ is a small threshold. 

The objective is approximated using a first-order Taylor expansion:
\[
\mathbb{E}_{\tau \sim \pi_\theta} \left[ \frac{\pi_{\theta'}(a_t | s_t)}{\pi_\theta(a_t | s_t)} A^{\pi_\theta}(s_t, a_t) \right] \approx g^T (\theta' - \theta),
\]
where $g = \nabla_\theta J(\theta)$ is the policy gradient.

The KL divergence constraint is approximated using a second-order Taylor expansion:
\[
\mathbb{E}_{s \sim \pi_\theta} \left[ D_{\text{KL}} \left( \pi_\theta(\cdot|s) \| \pi_{\theta'}(\cdot|s) \right) \right] \approx \frac{1}{2} (\theta' - \theta)^T H (\theta' - \theta),
\]
where $H$ is the Fisher Information Matrix.

The constrained optimization problem becomes:
\[
\max_{\theta'} g^T (\theta' - \theta) \quad \text{subject to} \quad \frac{1}{2} (\theta' - \theta)^T H (\theta' - \theta) \leq \delta.
\]

The solution, using Lagrange multipliers, is:
\[
\theta' = \theta + \sqrt{\frac{2\delta}{g^T H^{-1} g}} H^{-1} g.
\]

This step ensures that the update direction is the natural gradient, scaled to satisfy the trust region constraint:
\[
\theta_{k+1} = \theta_k + \alpha H^{-1} g.
\]

If we were to stop here, the algorithm would compute the Natural Policy Gradient. However, to ensure that the KL constraint is satisfied, TRPO introduces a backtracking line search. The update rule becomes:
\[
\theta_{k+1} = \theta_k + \alpha^j \sqrt{\frac{2\delta}{g^T H^{-1} g}} H^{-1} g,
\]
where \( \alpha \in (0, 1) \) is the backtracking coefficient, and \( j \) is the smallest non-negative integer such that \( \pi_{\theta_{k+1}} \) satisfies the KL constraint and improves the objective.

\subsection{Proximal Policy Optimization (PPO)}

PPO improves upon TRPO by simplifying the trust region constraint. PPO introduces a clipped surrogate objective to penalize overly large changes to the policy. Let \( r_t(\theta) \) be the probability ratio:
\[
r_t(\theta) = \frac{\pi_\theta(a_t | s_t)}{\pi_{\theta_{\text{old}}}(a_t | s_t)},
\]
so \( r(\theta_{\text{old}}) = 1 \). Recall that TRPO maximizes the surrogate objective:
\[
L^{TRPO}(\theta) = \mathbb{E}_t \left[ r_t(\theta) \hat{A}_t \right].
\]
However, maximizing \( L^{CPI}(\theta) \) without constraints can lead to excessively large policy updates. PPO modifies this with the following objective:
\[
L^{CLIP}(\theta) = \mathbb{E}_t \left[ \min \left( r_t(\theta) \hat{A}_t, \, \text{clip} \left( r_t(\theta), 1 - \epsilon, 1 + \epsilon \right) \hat{A}_t \right) \right],
\]
where \( \epsilon \) is a hyperparameter (often \( \epsilon = 0.2 \)) that controls the range for clipping. The clipping of \( r_t(\theta) \) ensures that the policy does not change too much by capping the probability ratio within \( [1 - \epsilon, 1 + \epsilon] \).

\section{Fictitious Play Algorithms}

\subsection{Extensive-Form Games}
An extensive-form game is represented by a game tree with the following components:
\begin{itemize}
    \item \( N = \{1, \dots, n\} \): Set of players.
    \item \( S \): Set of states, each corresponding to a decision node in the tree.
    \item \( A(s) \): Set of actions available at state \( s \in S \).
    \item \( P: S \to N \cup \{c\} \): Player function, where \( c \) denotes chance.
    \item \( U_i \): Information sets for player \( i \), denoting indistinguishable states.
    \item \( \pi_i(u) \in \Delta(A(u)) \): Player \( i \)'s behavioral strategy at information set \( u \).
    \item \( R(s) \): Payoff function at terminal states.
\end{itemize}

\subsection{Fictitious Play (FP)}
Fictitious play (Brown, 1951) is defined as normal-form games. Each player repeatedly plays a best response to the empirical frequency of opponents' past actions, updating their strategy over time.

At each time step \( t \), player \( i \) chooses the best response \( \beta_i^t \) to the empirical distribution of their opponents' strategies up to time \( t \). Let \( \sigma_{-i}^t \) denote the empirical mixed strategy of all players except player \( i \) at time \( t \). The best response for player \( i \) is then:
\[
\beta_i^t = \arg\max_{a_i \in A_i} \mathbb{E}_{\sigma_{-i}^t}[u_i(a_i, \sigma_{-i}^t)],
\]
where \( \mathbb{E}_{\sigma_{-i}^t} \) denotes the expectation over the mixed strategy \( \sigma_{-i}^t \) of the opponents.

Opponents' beliefs about the empirical strategy \( \sigma_i^{t+1} \) for player \( i \) at time \( t+1 \) are updated as:
\[
\sigma_i^{t+1}(a_k) = \frac{1}{t+1} \sum_{\tau=1}^{t} \mathbb{1}(\beta_i^\tau=a_k).
\]

\subsection{Weakened Fictitious Play}

Weakened fictitious play (Van der Genugten, 2000) is identical to fictitious play, except that at each step the strategies played need only be \(\epsilon\)-best responses. An \(\epsilon\)-best response for player \( i \) to the opponents' strategy profile \(\sigma_{-i}\) is defined as:
\[
\beta_i^{\epsilon}(\sigma_{-i}) = \left\{ \sigma_i \in \Delta_i : u_i(\sigma_i, \sigma_{-i}) \geq u_i(\beta_i(\sigma_{-i}), \sigma_{-i}) - \epsilon \right\},
\]
where \( u_i \) is the expected payoff for player \( i \), and \( \beta_i(\sigma_{-i}) \) is the true best response. An \(\epsilon\)-best response is a strategy that performs no more than \(\epsilon\) worse than the best response.

\subsection{Generalized Weakened Fictitious Play}

Generalized Weakened Fictitious Play (Leslie \& Collins, 2006) extends weakened fictitious play by allowing approximate best responses with small perturbations in strategy updates:
\[
\sigma_i^{t+1}(a_k) = (1 - \alpha_{t+1}) \sigma_i^t(a_k) + \alpha_{t+1} \left( \beta_i^{\epsilon_t}(a_k, \sigma_{-i}^t) + M_i^{t+1}(a_k) \right),
\]
where:
\begin{itemize}
    \item \( \sigma_i^t(a_k) \) is player \( i \)'s mixed strategy at time \( t \),
    \item \( \alpha_{t+1} \to 0 \) is a decaying learning rate,
    \item \( \epsilon_t \to 0 \) controls how closely the best response approximates the optimal strategy,
    \item \( \beta_i^{\epsilon_t}(a_k, \sigma_{-i}^t) \) is the approximate best response,
    \item \( M_i^{t+1}(a_k) \) is a perturbation term introducing stochasticity.
\end{itemize}
This framework allows for smoother convergence and captures bounded rationality through decaying learning rates and suboptimal responses.

\subsection{Extensive Width Fictitious Play (Hendon et al., 1996)}
Updating strategies over the entire action space can be computationally expensive in normal-form games. **Extensive width fictitious play** focuses on behavioral strategies and operates on information sets. It allows more efficient updates, scaling with the number of information sets rather than the full action space.

For a player \( i \), the behavioral strategy update is:
\[
\pi_i^{t+1}(u) = \pi_i^t(u) + \alpha_{t+1} x_{\beta_i^{t+1}}(\sigma_u) \left( \beta_i^{t+1}(u) - \pi_i^t(u) \right),
\]
where:
\begin{itemize}
    \item \( u \) is the information set,
    \item \( \beta_i^{t+1}(u) \) is the best response strategy,
    \item \( x_{\beta_i^{t+1}}(\sigma_u) \) is the realization probability of reaching \( u \) given strategy \( \beta_i^{t+1} \),
    \item \( \alpha_t \) is the learning rate.
\end{itemize}

This update mechanism operates on the tree's information sets and adjusts strategies locally, resulting in improved computational performance for large extensive-form games.

\subsection{Neural Fictitious Self-Play (NFSP)}
Neural Fictitious Self-Play (NFSP) is an advanced algorithm combining reinforcement learning (RL) with fictitious play. It uses two components:
\begin{enumerate}
    \item \textbf{Reinforcement Learning Component (Best Response)}: Learns the best response to opponents using Q-learning. The Q-values are updated as:
    \[
    Q(s_t, a_t) \leftarrow Q(s_t, a_t) + \alpha \left[ r_t + \gamma \max_{a'} Q(s_{t+1}, a') - Q(s_t, a_t) \right],
    \]
    where \( \alpha \) is the learning rate and \( \gamma \) is the discount factor.
    
    \item \textbf{Supervised Learning Component (Average Strategy)}: Imitates the agent's own behavior by learning an average strategy from replayed data. The average strategy is updated by minimizing the cross-entropy loss:
    \[
    \mathcal{L}(\theta) = - \mathbb{E}_{(s, a) \sim \mathcal{D}} \log \pi(a|s; \theta),
    \]
    where \( \mathcal{D} \) is the replay memory containing state-action pairs and \( \pi(a|s; \theta) \) is the policy parameterized by \( \theta \).
\end{enumerate}

The NFSP algorithm alternates between these two components, where the reinforcement learning component continuously improves the best response, and the supervised learning component builds a stable average strategy. This results in convergence toward Nash equilibria in multi-agent environments.

\begin{algorithm}[H]
\caption{Neural Fictitious Self-Play (NFSP)}
\textbf{Initialize}: Replay memories \( \mathcal{M}_{RL} \), \( \mathcal{M}_{SL} \); Q-network \( Q(s, a \, | \, \theta^Q) \); Policy network \( \pi(s, a \, | \, \theta^\pi) \)\;
\For{each episode}{
    \For{each agent}{
        Observe state \( s_t \)\;
        Sample action \( a_t \sim \epsilon\)-greedy\;
        Execute action and observe reward \( r_t \) and next state \( s_{t+1} \)\;
        Store transition \( (s_t, a_t, r_t, s_{t+1}) \) in \( \mathcal{M}_{RL} \)\;
        Update Q-network by minimizing:
        \[
        \mathcal{L}(\theta^Q) = \left( r_t + \gamma \max_{a'} Q(s_{t+1}, a'; \theta^Q) - Q(s_t, a_t; \theta^Q) \right)^2
        \]
        Store action \( (s_t, a_t) \) in \( \mathcal{M}_{SL} \)\;
    }
    Update policy network by minimizing the cross-entropy:
    \[
    \mathcal{L}(\theta^\pi) = - \log \pi(a_t|s_t; \theta^\pi)
    \]
}
\end{algorithm}

This combined approach of reinforcement learning and supervised learning enables agents to converge to Nash-like behavior over time, even in complex, imperfect information games.

\end{document}